\documentstyle[aps,epsf]{revtex}     
\newcommand{\postscript}[2] {\setlength{\epsfxsize}{#2\hsize}
\centerline{\epsfbox{#1}}}
\begin{document}
\title{Universal aspects of Efimov states and light halo nuclei}
\author{A. E. A. Amorim$^{(a,b)}$, T. Frederico$^{(c)}$ and 
Lauro Tomio$^{(a)}$}
\address{
$^{(a)}$ Instituto de F\'\i sica Te\'orica, Universidade Estadual Paulista, 
01405-900 S\~{a}o Paulo, Brasil \\
$^{(b)}$ Faculdade de Tecnologia de Jahu, CEETEPS,
Jahu, Brasil \\ 
$^{(c)}$ Dep. de F\'\i sica, 
Instituto Tecnol\'ogico de Aeron\'autica, Centro T\'ecnico Aeroespacial,\\
	    12228-900 S\~ao Jos\'e dos Campos, Brasil }
\vskip 1cm
\maketitle
\begin{abstract}
The parametric region in the plane defined by the  ratios of the energies 
of the subsystems and the three body ground state, in which Efimov states 
can exist, is determined. 
We use a renormalizable model that guarantees the general validity 
of our results in the context of short range interactions.
The experimental data for one and two-neutron separation 
energies, implies that among the halo-nuclei candidates, only $^{20}C$ 
 has a possible Efimov state, with energy less than 14 KeV 
below the scattering threshold.
\end{abstract}
%\pacs{PACS 21.10.Dr, 21.45.+v, 03.65.Ge}
\vskip 0.5cm

The Efimov effect\cite{ef70} is the remarkable theoretical observation that
the number of bound states for three particles interacting via s-wave 
short range potentials may grow to infinity, as the pair interactions are just 
about to bind two particles. Such Efimov states are loosely bound and their 
wave functions extend far beyond those of normal states.
If such states exist in nature they 
will dominate the low-energy scattering of one of the particles with the
bound-state of the remaining two particles. Such states have been studied in
several numerical model calculations\cite{efim2,tom,fe93,realist}. 
There were searches 
for Efimov states in atomic and nuclear systems without a clear experimental 
evidence of occurrence of such states\cite{evid1,ef90}. 

The discovery of halo nuclei brought a lot of attention to the 
search for Efimov states, because such systems can be viewed as a three body 
system with two loosely bound neutrons and a core which is more bound
\cite{tani,hu93,zhu93}.  
Fedorov, Jensen and Riisager\cite{realist}, in a first calculation of possible 
Efimov states in halo nuclei, suggested as promising candidates the nuclei 
$^{18}C$ and $^{20}C$. They also suggest other possible occurrence in 
neutron rich oxygen isotopes.

The light three-body 
halo nuclei have unusual properties in respect to the nuclear size; 
the radius of the neutron halo is much greater than the radius of the core,
and the core is assumed structureless \cite{fe93,realist,das94}. 
For the moment, we use this assumption and later on, we discuss
its validity in our calculation.
This situation allows the use of concepts coming from short-range
interactions. 

In this Letter, we settle the general basis for the existence of 
Efimov states, through  the use of universal properties of three body 
systems at low energies, in a way convenient to analyze weakly bound 
systems like the halo nuclei. The approach is parameterized
by the two and three body energies in a zero-range model. 
The renormalizability of the quantum mechanical many-body model with the 
s-wave zero-range force, 
implies that all the low-energy properties of the three-body system
are well defined if one three-body and  the low-energy two body physical 
informations are known \cite{ad95}. The three body input
can be chosen as the experimental ground state binding energy. All the 
detailed informations about the short-range force, beyond 
the low-energy two-body 
observables, are retained in only one three body physical information in the 
limit of  zero range interaction. 
The sensibility of the three body binding energy to the interaction
properties comes from the collapse of the system in the 
limit of zero-range force, the Thomas effect \cite{th35}. 

The domain of the coupling constants to guarantee three or four body 
bound states, when the subsystems are unbound, was studied in 
Ref.~\cite{ri94}, using short-range interactions. 
In our approach instead of the strength of the interaction, a quantity not 
directly available, we fix the energies of the subsystems (bound or virtual) 
and the three-body ground state by the available experimental data 
and look for the excited states. 

As the input energies  are fixed in the renormalized model, 
a more realistic potential will not
affect the generality of the present conclusions.
In that sense, the Pauli principle correction, between the halo and 
the core neutrons, affects essentially the ground state energy which is 
already fixed. We have to consider that this is a short-range phenomenon 
that occurs for distances less than the core size (about $\approx 3 fm$ 
for light-halo nuclei). 
We believe that our results are valid even in the case where the spin of the 
core is non zero.
Considering that we give as input the experimental energies of the 
ground-state three 
body system and the sub-systems,  the  spin 
effect is already being considered by this procedure. 
Also, theoretical evidence was found in the last reference of \cite{fe93}, 
that the inclusion of core spin  does not alter significantly the 
possibility to have an Efimov state.

The notation used is appropriate for halo nuclei, $n$ for neutron
and $c$ for core, but  our approach 
is applicable  to any three-particle system which interact 
via s-wave short-range interactions,  where two of the particles 
are identical.  The s-wave interaction for the $n-c$ potential is 
justified because the excited state (if exists)
should have an  extremely small energy, just allowing zero angular 
momentum for the two-particle  state in the relative coordinates 
\cite{fe93}.
It was already observed in Ref.\cite{zin}, when discussing $^{11}Li$, that 
even the three-body wave-function with an s-wave $n-n$ correlation 
produces a ground  state of the halo nuclei with
two or more shell-model configurations.  

The energies of the two particle subsystem, $E_{nn}$ and $E_{nc}$ can be 
virtual or bound. In a plane defined by these two observables, for example,
the $^{11}Li$ will be represented by a point given by the well known
virtual-state of the neutron-neutron subsystem (143 KeV), and the not so well 
determined low-energy virtual state of $^{10}Li$ ($\sim$ 50 KeV,  
according to \cite{zin}). We can vary
the mass of the core to hold several other low 
masses halo nuclei, like $^{11}Li$, $^{12}Be$, $^{18}C$ and $^{20}C$.

We can anticipate the qualitative behaviour of the first Efimov excited state
in the halo nuclei by changing the $n-n$ and $n-c$ energies.
The Efimov states should disappear in the two-body threshold with the
increase of the two-body binding energy \cite{tom}.  Another result  obtained
in a previous study shows that the variation of the three-body energy around
a vanishing two-body energy is proportional to the square of the 
absolute value of
the two-body bound or virtual state energy \cite{am92}. The three-body energy
increases with the  two-body bound state energy and decreases in the
direction of a two-body virtual state. As we will see in the following, this
corresponds to  weaken the kernel of three-body zero-range equations for the
halo-nuclei. Comparing with Ref. \cite{ri94},
in our discussion the weakening of the strength is simulated by  
increasing the two-body virtual state energy.

The zero-range three-body integral equations for the bound state of two
identical particles and a core, is written as 
generalization of the three-boson equation \cite{st57}. 
It is composed by two coupled integral equations  close to the
s-wave separable potential model of Ref.\cite{das94}. 
The antisymmetrization of the two outer neutrons 
is satisfied since the spin couples to zero \cite{fe93}.
In our approach the potential form factors and
corresponding strengths are replaced, in the
renormalization procedure, by the two-body binding energies, 
$E_{nn}$ and $E_{nc}$. 
In the case of bound systems, these quantities are the separation energies.
We distinguish these two cases  by the following definition:  
$ K_{nn} \equiv \pm \sqrt{E_{nn}} , \; \; \; 
K_{nc} \equiv \pm \sqrt{E_{nc}} $ , 
where $+$ refers to bound and $-$ to virtual state energies.
Our units will be such that $\hbar = 1$ and nucleon mass $=1$.

After partial wave projection, the s-wave 
$n-n-c$ coupled integral equations are:
\begin{eqnarray}
\chi_{nn}(q)&=&\tau_{nn}(q;B_N)
\int_0^\Lambda dk G_1(q,k;B_N)\chi_{nc}(k) 
\label{chi1} \\ 
\chi_{nc}(q)&=&\tau_{nc}(q;B_N)\int_0^\Lambda dk 
\left[G_1(k,q;B_N) \chi_{nn}(k) 
+ A G_2(q,k;B_N) \chi_{nc}(k)\right] ,
\label{chi2} 
\end{eqnarray}
where
\begin{eqnarray}
\tau_{nn}(q;E)&=&
\frac{2}{\pi}
\left[\sqrt{E+\frac{A+2}{4A} q^2}-K_{nn}\right]^{-1},
\label{tau1} 
\end{eqnarray}
\begin{eqnarray}
\tau_{nc}(q;E)&=&\frac{1}{\pi}\left(\frac{A+1}{2A}\right)^{3/2}
\left[\sqrt{E+\frac{A+2}{2(A+1)} q^2}-K_{nc}\right]^{-1},
\label{tau2}
\end{eqnarray}
\begin{eqnarray}
G_1(q,k;E)&=&\frac{k}{q}\log 
\frac{2A(E+k^2+qk)+q^2(A+1)}{2A(E+k^2-qk)+q^2(A+1)},
\label{G1} 
\end{eqnarray}
\begin{eqnarray}
G_2(q,k;E)&=&\frac{k}{q} \log 
\frac{2(A E+qk)+(q^2+k^2)(A+1)}{2(A E-qk)+(q^2+k^2)(A+1)}.
\label{G2} 
\end{eqnarray}
$B_N$ is the $N$-th three-body halo state energy
and $A$ is the core mass number.  The cut-off, $\Lambda$, 
represents the inverse of the interaction radius \cite{ef90} and it goes
to infinity as the radius of the interaction decreases.  We made the
assumption that the range of $n-n$ and $n-c$ potentials are about the same,
represented by just one radius.  This assumption should not be regarded as a
limitation, since the three-body model is renormalized,  
requiring only one three-body observable to be fixed\cite{ad95},
together with the two-body low-energy physical informations. The results are
independent of the regularization scheme, which in this case means different
cut-off's.

The Thomas collapse of the three-body binding energy is seen for
$\Lambda\rightarrow \infty$ in Eqs. (\ref{chi1}) and (\ref{chi2}). 
This limit is
equivalent to the situation that allows the Efimov states,
once Thomas and Efimov states are related by a scale 
transformation \cite{efth}.
To illustrate this connection, in this 
particular case, we can make a transformation of Eqs. (\ref{chi1} - \ref{G2}) 
to the units in which the  cut-off is one. The corresponding equations
are formally the same as given in Eqs. (\ref{chi1} - \ref{G2}), 
with the dimensional variables and observables replaced by the 
corresponding non-dimensional quantities, such that
$$\Lambda \to 1, \; B_N \to \epsilon_N
\equiv {B_N}/{\Lambda^2}, \;  
K_{nn} \to \kappa_{nn}\equiv {K_{nn}}/{\Lambda}, \;
K_{nc} \to \kappa_{nc}\equiv {K_{nc}}/{\Lambda}.$$
The two-body observables can be written
in terms of the three-body energy $B_N$, by replacing $\Lambda$,
such that
$${\kappa_{nn}}/{\sqrt{\epsilon_N}} = {K_{nn}}/{\sqrt{B_N}}\;
{\rm and} \; {\kappa_{nc}}/{\sqrt{\epsilon_N}} = {K_{nc}}/{\sqrt{B_N}}.$$
The collapse of the three-body energy and the presence of Efimov states
are consequences of the existence of solutions of the corresponding
equations in the limit of 
$\kappa_{nn} \rightarrow 0$ and $\kappa_{nc} \rightarrow 0$. 
The Thomas effect is seen
for $\Lambda\rightarrow \infty $ with the energies
of the two-body systems fixed, and the Efimov states are seen
for $K_{nn}\ \rightarrow 0 $ with
 $K_{nc} \ \rightarrow 0$ and $\Lambda$ fixed. 

The three-body halo energy scales with the cut-off
parameter and the Thomas collapse is seen by
the unbound increasing of the energies of the bound states 
 for $\Lambda\to\infty$. The values of $\epsilon_N$ in this 
limit for the first three states are given in Table I.
The excited state energies in the units where $\Lambda=1$ approaches
zero by increasing $N$. 
\vskip 1cm
\noindent {\small  {\bf Table I} \ 
First binding energies, in non-dimensional units 
as defined in the text, as a function of the core mass-number {\it(A)} of 
the collapsed three-body system.}
\begin{center}
\begin{tabular}{ccccccc}
\hline\hline
 $A =$          & 1           & 5           & 9           & 18     
& 50          & 100 \\
\hline 
$10^2\epsilon_0 =$&3.17\  &4.83\ &5.17\ & 5.41\ & 5.58\ & 5.68\  \\
$10^4\epsilon_1 =$&0.60\  &2.31\ &2.93\ & 3.46\ & 3.88\ & 4.15\  \\
$10^6\epsilon_2 =$&0.11\  &1.16\ &1.77\ & 2.38\ & 2.92\ & 3.29\  \\
\hline\hline
\end{tabular}
\end{center}
\vskip 1cm

In Fig.1 we have our main results. We show the parametric plane 
defined by $K_{nc}/ \sqrt{B_N}$ versus $ K_{nn}/ \sqrt{B_N}$.
The curves represent  
the boundary of the region where there is an excited $(N+1)$-th state above 
the $N$-th state, for $A$=1, 9, 18 and 100. 
The boundary curve means that the $(N+1)$ three-body binding energy is equal 
to the lowest scattering threshold. 
Outside such region the excited bound $(N+1)$ state does not exist. 
We found that the boundary is practically the same for $N=0$ and $N=1$.
The limiting boundary, for $N\rightarrow \infty $, 
corresponds to the renormalized result in the 
limit of the zero-range force.
In each step the $N$-th state plays the role of the
ground state and $(N+1)$-th state corresponds to the first
excited state on the top of the $N$-th state. 
The energy scale of the $n-n-$core system is given by $B_N$.
The fast approach of   
the  limiting boundary with $N$ is due to the numerical 
value of $\epsilon_0$, much smaller than 1, as seen in Table I.

We show four different regions in Fig.1,
where the free two-body  subsystems have a virtual or bound
state. In region (I), the $n-n$ and $n-c$
subsystems  have both bound states,  and in such case
$\kappa_{nn}$ and $\kappa_{nc}$ are positive. 
In region (II), the $nc$-subsystem is bound and 
the $nn$-subsystem has a virtual state, such that 
$K_{nc}>0$ and $K_{nn}<0$.
The region (III), where $K_{nn}<0$ and $K_{nc}<0$, both 
subsystems $n-n$ and $n-c$ have virtual states. 
The region (IV) has $K_{nc} < 0$ and $K_{nn}>0$. 

We observe an asymmetry between regions (II) and (IV), that can be explained
due to the fact that we have two interactions of the kind $n-c$ (between the
particle $n$ and the core $c$), and just one kind of $n-n$  interaction.  So,
if $n-n$ is  virtual and 
$n-c$ is bound, as in region (II), we have two possibilities of two-body
bound states with just one possibility of a two-body virtual state. Therefore,
we can easily see that the room for a three-body bound state is very much
reduced in region (IV) compared with region (II), as in region (IV) only one
third of the two-body interactions is giving favorable conditions for
binding.

Using this plot we can analyze the existence of Efimov states in any 
three body system that interacts via attractive 
short range potentials, once the 
three body ground state energy and  
the virtual or bound two body energies are known.

\postscript{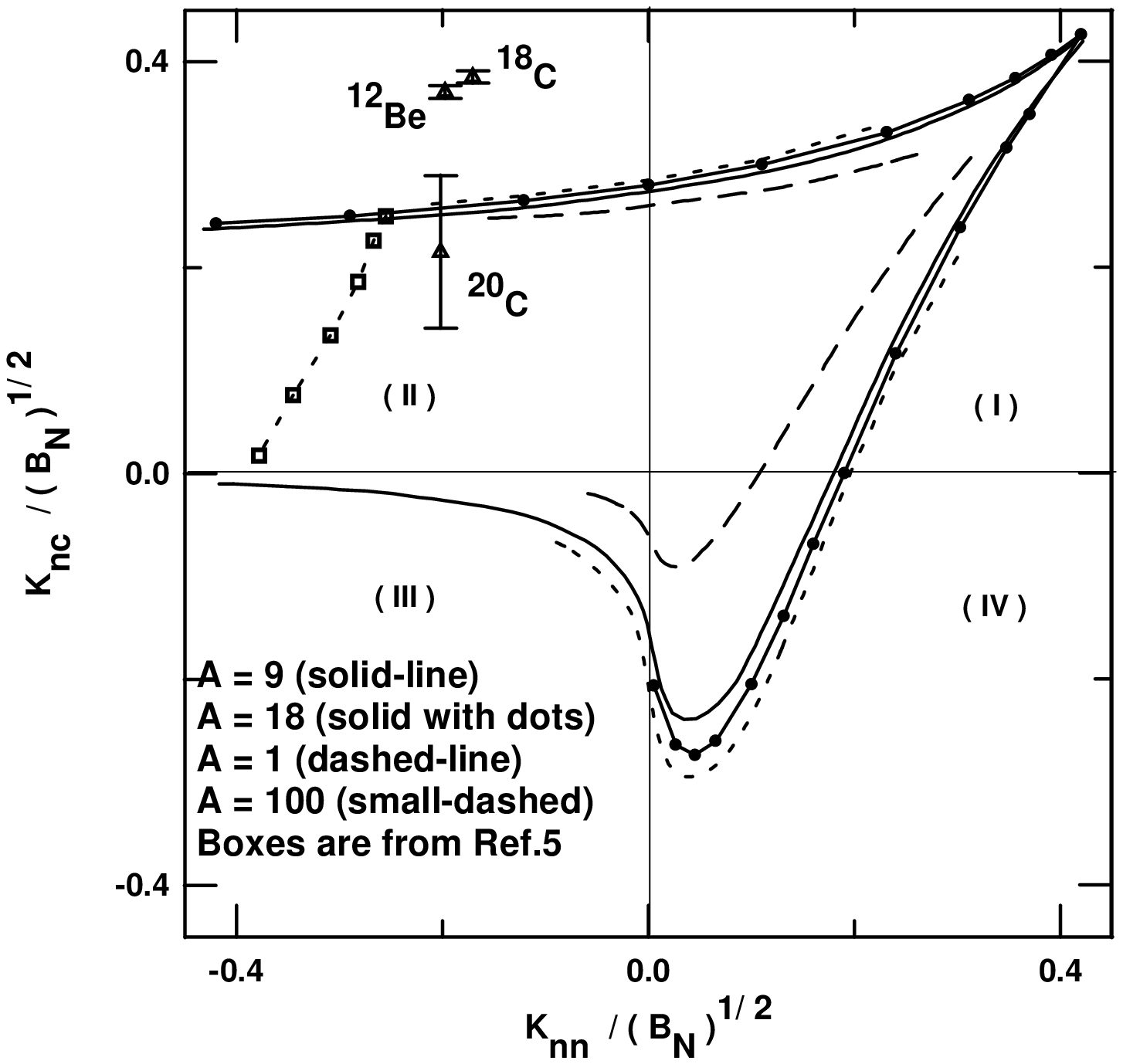}{0.6}    %**epsf**
\vskip -2cm
\noindent {\small 
{\bf Fig.1} \ \  
$K_{nn}/{(B_N)}^{1/2}$ as a function of  $K_{nc}/{(B_N)}^{1/2}$ at 
the boundary of the region where the binding energy of the $ (N+1) $-th 
Efimov state is zero. Negative values for the two body observables 
correspond to virtual states. Boundary curves for different core masses 
($A$); solid-line  for $A=9$; line with black dots for $A=18$; 
small-dashed-line for $A=100$ and dashed-line for $A=1$. \ We also show 
three experimental data, corresponding to the halo nuclei $^{20}C$, 
$^{18}C$ and $^{12}Be$ [19].
The squares, connected with dashed lines, are obtained from Fig.2 of 
Ref. [5] for $A=9$.
}
\vskip 1cm

For the halo nuclei, the relevant
section of the plot is the regions II and III, because 
the $n-n$ subsystem has a virtual state ($E_{nn}$=143 KeV). 
The region we show can accommodate only 
a few halo nuclei and only  the experimental data for $^{20}C$ \cite{audi}
is practically inside it the area delimited by the curve that 
allows Efimov states. Considering the small experimental
errors in the data for $^{12}Be$, $^{18}C$ \cite{audi},
we cannot expect to find Efimov states in such nuclei. 
We are not showing the point corresponding to $^{11}Li$, because the 
actual value for the virtual state of $^{10}Li$\cite{zin} ($\sim 50$ KeV)
is ruling out the existence of Efimov states in this nucleus. Only if
$^{10}Li$ had an almost zero virtual energy or being bound with
an energy less than $\sim 15$ KeV, it would 
be possible an Efimov state.
The results for $^{11}Li$ with a bound $^{10}Li$
given in Fig.2 of Ref.\cite{realist},  are completely inside  region II,
with an exact agreement of the corresponding threshold. 

\postscript{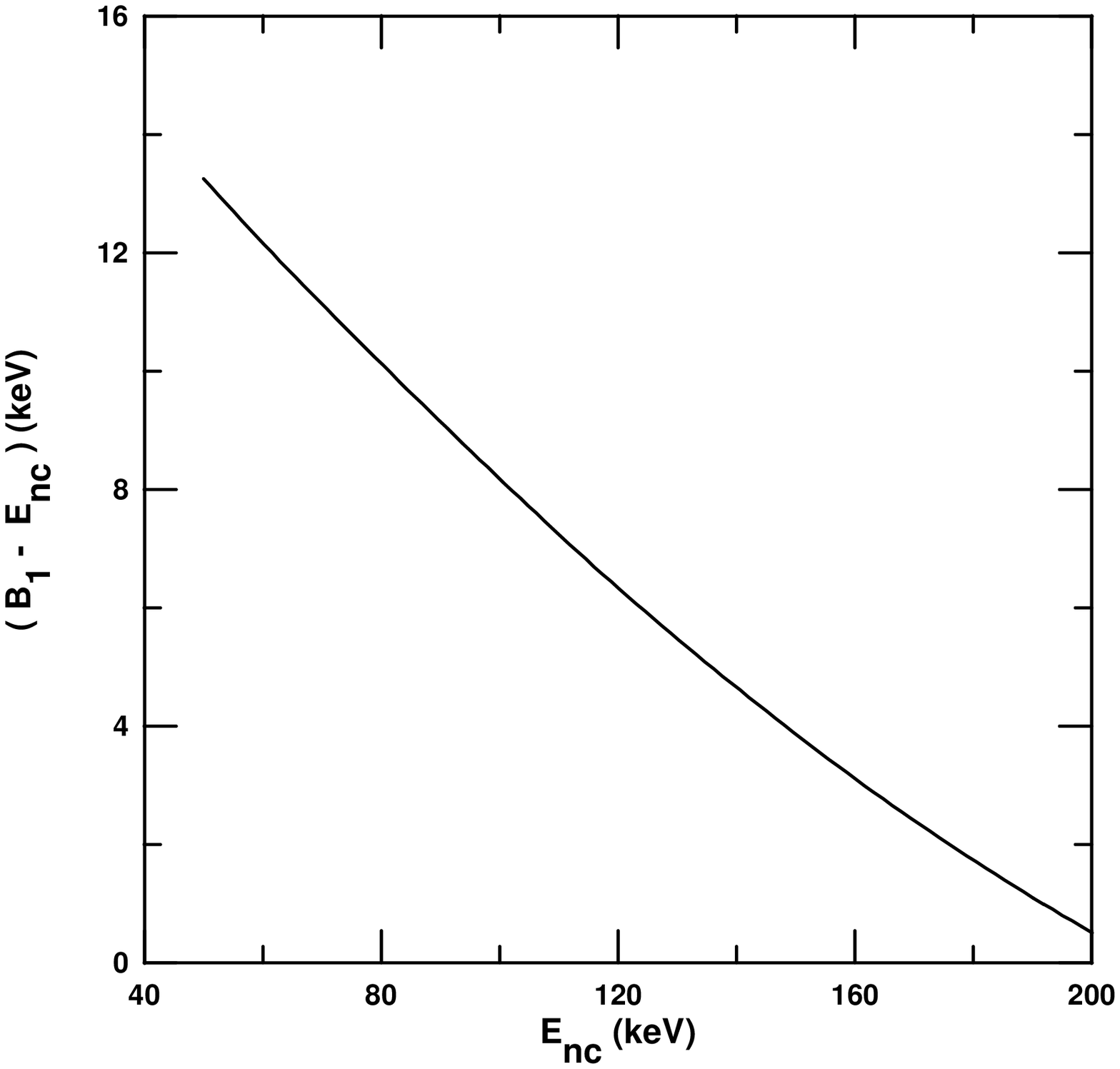}{0.5}    %**epsf**
\vskip -1.5cm
\noindent {\small {\bf Fig. 2} \ \
Binding energy of the $^{20}C$ Efimov state relative to the elastic 
threshold as a function of the $^{19}C$ neutron bound state energy. 
}
\vskip 1cm

A brief discussion of the applicability of the results to halo nuclei is
needed. The many-body structure of the $N$'th state 
manifests in the n-n-core system, when the two neutrons are inside the core.
The three body dynamics is effective for relative distances
bigger than the core size and interaction radius ($r_0$), if
$a_{nn}\ >> \ r_0$  and $a_{nc}\ >> \ r_0$ \cite{ef70,ef90},
with $a_{nn}$  and $a_{nc}$ being the
scattering lengths of the n-n and  n-core subsystems, respectively.
Even if the $N$'th state is rather bound, at separation distances where the 
three body dynamics is effective the  three-body wave-function tail is 
built with the knowledge of the binding energy. 
The three-body dynamics develops
a long range potential in the hyper-radius of the
three-body system \cite{ef70,ef90}, which carries the
informations about $a_{nn}$  and $a_{nc}$.
The excited state appears in the long range 
potential; it has the same angular components as  the 
ground state, and it is determined by the logarithmic derivative 
of the ground state wave-function at some hyper-radius ($r_b$),  such that
$r_0<r_b<<a$. This discussion supplies
the physical picture of the renormalized three-body model.
The values of $a_{nn}$ and $a_{nc}$  for 
$^{11}Li$, $^{12}Be$, $^{18}C$ and $^{20}C$ fulfill the condition
of being much bigger than the core size and interaction radius.
The systematic uncertainties in the application
can appear as powers of $r_0/a_{nn}$ and $r_0/a_{nc}$ \cite{ef90}
($\approx$ 0.1 for $^{20}C$). 

The experimental value of the separation energy of the neutron in 
$^{19}C$ has a sizeable error,
$E_{nc}=162\pm 112$KeV\cite{audi}, which allows a range for the excited state
energy. The experimental value of the three-body ground-state energy of
$^{20}C$ has a small error compared to the $E_{nc}$. Thus, we fixed the
ground state energy at $3506$KeV \cite{audi}
and study  the excited state energy as a function of
$E_{nc}$, as given in Fig. 2. We observe that if $E_{nc} >$ 200
KeV, the excited state is destroyed, but in the region of 50 KeV $< E_{nc} <$
200 KeV, the $^{20}C$ halo-nuclei supports the existence of an Efimov state.
The binding energy relative to the $n-(nc)$ scattering  threshold is below 14
KeV.  
We  estimate the size of this excited state to be at least 
35 fm, supporting its interpretation as an Efimov state.
The accuracy of the calculation of the binding energy of the Efimov 
state will depend mainly on the input experimental energies
(the relative error in the binding energy
of the neutron in $^{19}C$, is about 0.7).
Our calculation for $^{20}C$ relies on the experimental 
evidence \cite{ba} that the last neutron of $^{19}C$ is in
an intruder s-orbit.

Actually, the importance of core polarization in halo nuclei has been
debated in two recent works\cite{kr}. Our method is independent of 
effects that influence the
ground state energy of the system and/or the energies of the 
sub-systems, considering that such energies are given as inputs 
in the renormalizable three-body model. 
Although the  reference of Kuo, Krmpoti\'c and Tzeng
 \cite{kr} shows that core polarization is
suppressed in halo-nuclei, even if it were to be of some residual importance, 
in our model its effect  on the binding of the Efimov state, 
would be taken into account through the energy of the ground state,
which is given as input.

In summary, we have discussed the universal aspects of three-body halo nuclei
in the limit of a zero-range interaction.  We use the correlation of the
value of the ground state energy and the first excited Efimov state, and find
the set of values of the $n-n$ and $n-c$ energies that allows the existence
of at least one excited Efimov state.  Considering the available data,
we conclude that $^{20}C$ is the strongest candidate for having one excited 
Efimov state, with a binding energy below 14 KeV relative to the lowest 
scattering threshold. Our calculation and the available data exclude
the possibility of having Efimov states in any other light halo nuclei
known to us, like 
$^{11}Li$, $^{12}Be$, $^{18}C$ or oxygen isotopes. 

Our thanks for support from Funda\c c\~ao de Amparo \`a Pesquisa do Estado 
de S\~ao Paulo (FAPESP) and from Conselho Nacional de
Desenvolvimento Cient\'{\i}fico e Tecnol\'ogico (CNPq) of Brazil.


\begin{references}
\bibitem{ef70} V. Efimov, Phys. Lett. {\bf B33}, 563 (1970).
\bibitem{efim2} A. T. Stelbovics and L.R. Dodd, Phys. Lett. {\bf 39B}, 450
(1972); A.C. Antunes, V.L. Baltar and E.M. Ferreira, Nucl. Phys. {\bf A265},
365 (1976).
\bibitem{tom} S.K. Adhikari and L. Tomio, Phys. Rev. {\bf C 26}, 83 (1982);
S.K. Adhikari, A.C. Fonseca and L.Tomio, Phys. Rev. {\bf C 26}, 77 (1982).
\bibitem{fe93} D.V. Fedorov and A.S.  Jensen, Phys. Rev. Lett. {\bf 25}, 4103 
(1993); 
D.V. Fedorov, E. Garrido and A.S. Jensen, Phys. Rev. {\bf C 51}, 3052 (1995).
\bibitem{realist}D.V. Fedorov, A.S. Jensen and K. Riisager, Phys. Rev. Lett.
{\bf 73}, 2817 (1994). 
\bibitem{evid1} T.K. Lim, K. Duffy, and W.C. Damest, Phys. Rev. Lett.
{\bf 38}, 341 (1977); H.S. Huber, T.K. Lim and D.H. Feng, Phys. Rev. 
{\bf C 18}, 1534 (1978).
\bibitem{ef90} V. Efimov, Comm. Nucl. Part. Phys. {\bf 19}, 271 (1990);
and references therein.
\bibitem{tani} I. Tanihata, J.Phys. {\bf G22}, 157 (1996).
\bibitem{hu93} C.A.Bertulani, L.F. Canto and M.S.Hussein, Phys. Rep.  {\bf
226}, 281 (1993).
\bibitem{zhu93} 
P.G. Hansen, A.S. Jensen, and B. Jonson, Annu. Rev. Nucl. Part. Sci.
{\bf 45}, 591 (1995);
M.V. Zhukov, B.V. Danilin, D.V. Fedorov, J.M. Bang, I.S.
Thompson and J.S. Vaagen, Phys. Rep. {\bf 231}, 151 (1993).
\bibitem{das94} S. Dasgupta, I. Mazumdar and V.S. Bhasin,
Phys. Rev. {\bf C50}, 550 (1994).
\bibitem{ad95} S. K. Adhikari, T.Frederico and I.D. Goldman,
Phys. Rev. Lett. {\bf 74}, 487 (1995);S.K. Adhikari and T. Frederico
Phys. Rev. Lett. {\bf 74}, 4572(1995).
\bibitem{th35} L.H. Thomas, Phys. Rev. {\bf 47}, 903 (1935)
\bibitem{ri94}J.-M. Richard and S.Fleck, Phys. Rev. Lett. {\bf 73},
1464 (1994).
\bibitem{zin} M. Zinser et.al. Phys. Rev. Lett. {\bf 75}, 1719(1995).
\bibitem{am92}A.E.A. Amorim, L.Tomio and T.Frederico,
Phys. Rev. {\bf C46}, 2224 (1992).
\bibitem{st57} G.V. Skornyakov and K.A. Ter-Martirosian,
Sov. Phys. JETP {\bf 4}, 648 (1957).
\bibitem{efth}
 S. K. Adhikari, A. Delfino, T. Frederico, I. D. Goldman
 and L. Tomio,  Phys. Rev. {\bf A37}, 3666 (1988).
\bibitem{audi} G. Audi and A.H. Wapstra, Nucl. Phys. {\bf A595}, 409 
(1995); experimental values for the separation energies obtained
via ftp, as suggested in Ref.[1] of this reference.
\bibitem{ba}D.Bazin et al Phys. Rev. Lett.{\bf74}, 3569 (1995).
\bibitem{kr} T.T.S. Kuo, F. Krmpoti\'c, and Y. Tzeng,
Phys. Rev. Lett.{\bf 78}, 2708 (1997);
A. Cs\'ot\'o, "Importance of core polarization in halo nuclei",
nucl-th 9704054.
\end{references}
\end{document}